# Lattice swelling and modulus change in a helium-implanted tungsten alloy: X-ray micro-diffraction, surface acoustic wave measurements, and multiscale modelling

F Hofmann[1*], D Nguyen-Manh[2,3], CE Beck[3], JK Eliason[4], MR Gilbert[2], AA Maznev[4], W Liu[5], DEJ Armstrong[3], KA Nelson[4], SL Dudarev[2,3]

[1] Department of Engineering Science, University of Oxford, Parks Road, Oxford, OX1 3PJ, UK

[2] CCFE, Culham Science Centre, Abingdon, OX14 3DB, UK

[3] Department of Materials, University of Oxford, Parks Road, Oxford, OX1 3PJ, UK

[4] Department of Chemistry, Massachusetts Institute of Technology, 77 Massachusetts Avenue, Cambridge, MA 02139, USA

[5] Advanced Photon Source, Argonne National Lab, 9700 South Cass Avenue, Argonne, IL 60439, USA

* felix.hofmann@eng.ox.ac.uk

**Abstract:**

Using X-ray micro-diffraction and surface acoustic wave spectroscopy, we measure lattice swelling and elastic modulus changes in a W-1%Re alloy after implantation with 3110 appm of helium. A fraction of a percent observed lattice expansion gives rise to an order of magnitude larger reduction in the surface acoustic wave velocity. A multiscale elasticity, molecular dynamics, and density functional theory model is applied to the interpretation of observations. The measured lattice swelling is consistent with the relaxation volume of self-interstitial and helium-filled vacancy defects that dominate the helium-implanted material microstructure. Molecular dynamics simulations confirm the elasticity model for swelling. Elastic properties of the implanted surface layer also change due to defects. The reduction of surface acoustic wave velocity predicted by density functional theory calculations agrees remarkably well with experimental observations.

1. **Introduction:**

Tungsten (W) and tungsten-based alloys are the main candidate materials for plasma facing divertor surfaces in future fusion power plants [1] due to their high melting point, good resistance to sputtering, high thermal conductivity, and low tritium retention rate [2-5]. In addition to the radiation damage produced by collision cascades [6], bombardment of tungsten with 14.1 MeV fusion neutrons also transmutes it into other chemical elements through nuclear reactions. Calculations show that significant amounts of rhenium (Re) (~0.2 at. % year$^{-1}$), tantalum (~0.1 at. % year$^{-1}$) and osmium (on the order ~0.1 at. % year$^{-1}$) will be produced during the operation of a Demo reactor [7]. The amount of helium (He) produced by transmutation in tungsten is relatively small (between 0.1 and 10 appm year$^{-1}$) [7]. During operation tungsten surfaces in the divertor will also be exposed to large fluxes of hydrogen isotopes and helium ions with a broad spectrum of energies, resulting in a high heat flux of up to 10-15 MW/m$^2$ [8].

High-flux low-energy helium ion implantation causes significant modification of the material surface (e.g. formation of sponge-like structures, "fuzz" and bubbles) even at ion energies below the sputtering threshold (~100 eV) [9-11]. Similar effects have been observed at intermediate helium ion energies (30 keV) and temperatures from 1000 K to 1400 K [12, 13]. At high temperatures it is expected that the implanted helium atoms diffuse deeper into the bulk, affecting material behaviour due to the strong helium interaction with radiation defects [14]. For example, it has been found experimentally, by nano-indentation, that the combined effect of helium implantation and cascade damage from self-ion irradiation on hardness is far greater than that of cascade damage alone [15].

An important question concerns the dominant mechanism by which helium is retained in the tungsten matrix. Positron Annihilation Spectroscopy (PAS) studies [16, 17] and ab-initio calculations [18] indicate that helium-induced microstructure in metals is driven by the propensity of helium atoms to form bound complexes with vacancies. These may be pre-existing vacancies or vacancies formed by helium agglomeration and self-trapping leading to the spontaneous production of Frenkel pairs [19, 20]. Helium also has a significant effect on radiation-induced microstructure through the suppression of vacancy and self-interstitial atom recombination since helium atoms rapidly fill vacant lattice sites [21]. The resulting availability of excess self interstitial atoms (SIAs) stimulates nucleation and growth of interstitial dislocation loops at elevated temperatures [22].

Due to these complex interactions, carrying out quantitative analysis of the microstructure formed as a result of helium ion implantation proves challenging. The majority of recoil events produced by helium ions during implantation have energy lower than ~100 eV (see Fig. 1 (c)). Such low energy recoil events produce only individual Frenkel pairs, and hence the microstructure of helium ion irradiated tungsten is expected to be dominated by small helium-vacancy clusters and small clusters of self-interstitial ion defects. The objective of the analysis given below is to correlate experimentally measured strains and various elastic properties with the notion of microstructure dominated by small helium-vacancy and self-interstitial atom clusters, through the use of data on defect properties derived from *ab initio* calculations.

## 2. **Experimental Measurements:**

### 2.1. **Sample preparation:**

To mimic transmutation-induced production of rhenium in tungsten, a W – 1 at. % Re alloy was manufactured by arc melting from high purity elemental powders [23]. 1 mm thick slices were polished using diamond paste and 50 nm colloidal silica suspension to produce a flat, damage-free surface. Optical micrographs show equiaxed grains with sizes ranging from 100 to 1000 μm (Fig. 1 (a)). Electron back scattered diffraction (EBSD) indicated no significant texture.

Helium ions were implanted at 300 °C to a depth of ~2.8 μm using a 2 MeV accelerator at the National Ion Beam Centre, University of Surrey, UK. To achieve a near uniform helium ion concentration in excess of 3000 appm throughout the implanted layer, implantations were carried out at 12 different energies and fluences [23]. The resulting implantation profile, predicted by the Stopping and Range of Ions in Matter (SRIM) code [24], assuming a Frenkel pair formation threshold energy of 68 eV [25], is shown in Fig. 1 (b). At depths between 1 and 2 μm the calculated implanted helium dose is 3110 ± 270 appm. The associated displacement damage is 0.24 ± 0.02 displacements per atom (dpa). Analysis of recoils caused by the implanted helium ions (Fig. 1 (c)) shows that they are predominantly low energy events. Hence we expect the formation of individual Frenkel pairs during implantation, rather than clusters of defects [26]. This is also confirmed by transmission electron microscopy (TEM) of other tungsten samples implanted with helium under the same conditions, which showed no visible defects after implantation, indicating that all the implantation-induced defects are below the TEM resolution limit [15].

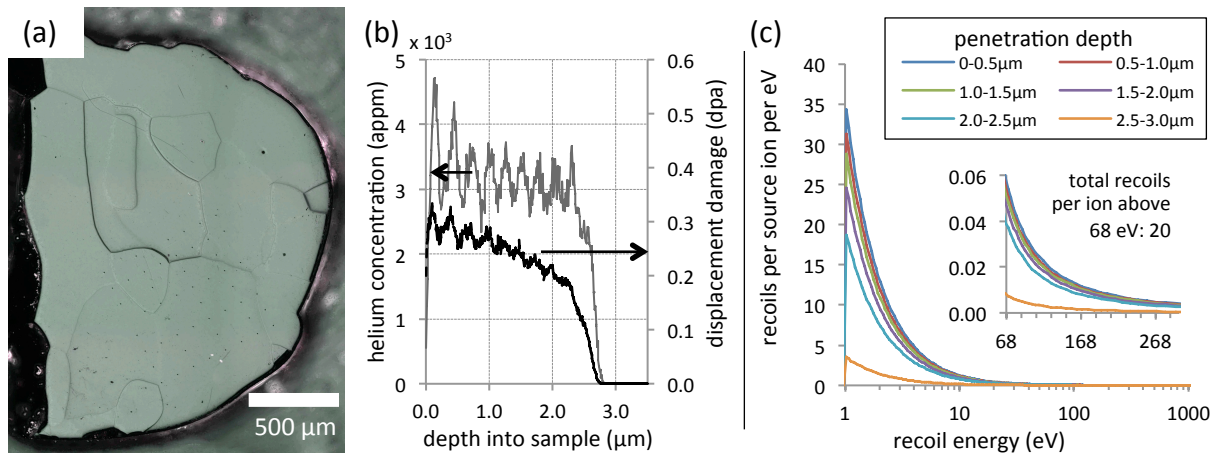

**Figure 1:** (a) Representative optical micrograph of the W – 1 % Re material. (b) SRIM calculated profile of injected helium ion concentration (grey curve) and implantation-induced displacement damage (black curve) as a function of depth in the sample. (c) Primary recoil atom energy spectrum per source ion calculated for different depths in the sample. On average each source ion produces 20 recoils with energy greater than the Frenkel pair formation threshold energy of 68 eV.

2.2. **Micro-diffraction measurements:**

Lattice swelling due to helium implantation was measured by micro-beam Laue diffraction at beamline 34-ID-E at the Advanced Photon Source, Argonne National Lab, USA. Fig. 2 (a) shows a schematic of the experimental setup. The incident, polychromatic (7 – 30 keV) X-ray beam was focussed by Kirkpatrick-Baez (KB) mirrors to a probe spot with 600 nm vertical and 400 nm horizontal full width at half maximum and near Lorentzian shape. The sample was positioned at the focus in 45° reflection geometry. Diffraction patterns were recorded on a Perkin Elmer flat panel detector mounted in 90° reflection geometry above the sample. The Differential Aperture X-ray Microscopy (DAXM) technique was used to determine the depth in the sample from which different scattered contributions originated [27-29]. For DAXM measurements a 50 μm diameter platinum wire was scanned through the diffracted beams. By triangulating using the wire edge, depth-resolved Laue patterns were reconstructed at 500 nm intervals along the incident beam direction. A detailed description of the experimental setup and analysis routines is provided elsewhere [28, 30, 31].

Laue diffraction patterns containing more than 20 peaks were collected at 3 positions in the implanted sample. The XMAS (https://sites.google.com/a/lbl.gov/bl12-3-2/user-resources) [32] and LaueGo (J.Z. Tischler: tischler@anl.gov) software packages were used to determine the deviatoric lattice strain tensor, $\boldsymbol{\varepsilon}^*$, from the Laue patterns. By scanning of the incident X-ray beam energy, using the beamline monochromator ($\Delta E / E \approx 10^{-4}$), the absolute lattice spacing of the (5,1,4) reflection was determined. Combining this with the deviatoric strain tensor, $\boldsymbol{\varepsilon}^*$, the full elastic lattice strain tensor, $\boldsymbol{\varepsilon}$, was calculated [33-35]. By applying the DAXM technique to both polychromatic and energy resolved scans the depth variation of the total lattice strain tensor, $\boldsymbol{\varepsilon}$, was measured. Fig.

2 (b) shows the $\varepsilon_{zz}$ out-of-plane strain component and the $\varepsilon_{xx}$ and $\varepsilon_{yy}$ in-plane strain components averaged over three measurement locations. The standard deviation of measured values (shown as error bars in Fig. 2 (b)) provides an indication of the experimental variation [30, 36, 37].

Elastic strains in the unimplanted substrate material, at depths greater than ~4 µm, are close to zero. The measured lattice constant in that region is 3.1650±0.0007 Å, close to the literature value of 3.16522±0.00009 Å for pure tungsten [38]. This provides a built-in validation for our strain measurements and suggests that 1 % rhenium alloying does not have a significant effect on the crystallography of the tungsten base.

In the helium-implanted surface layer there is a substantial increase in the $\varepsilon_{zz}$ out-of-plane strain component, whilst the $\varepsilon_{xx}$ and $\varepsilon_{yy}$ in-plane strain components remain close to zero. This suggests swelling of the crystal lattice due to helium implantation and irradiation defects with constraint in the directions parallel to the surface due to the requirement of continuity at the interface between the implanted layer and the substrate. A similar effect has been observed in $UO_2$ implanted with 60 keV He ions [39]. The depth of 3 µm (Fig. 2 (b)) up to which the helium-implantation-induced positive (i.e. tensile) $\varepsilon_{zz}$ strain extends agrees well with the maximum implantation depth of 2.8 µm predicted by SRIM. This provides independent experimental validation for the light ion penetration depth predicted by SRIM calculations [24].

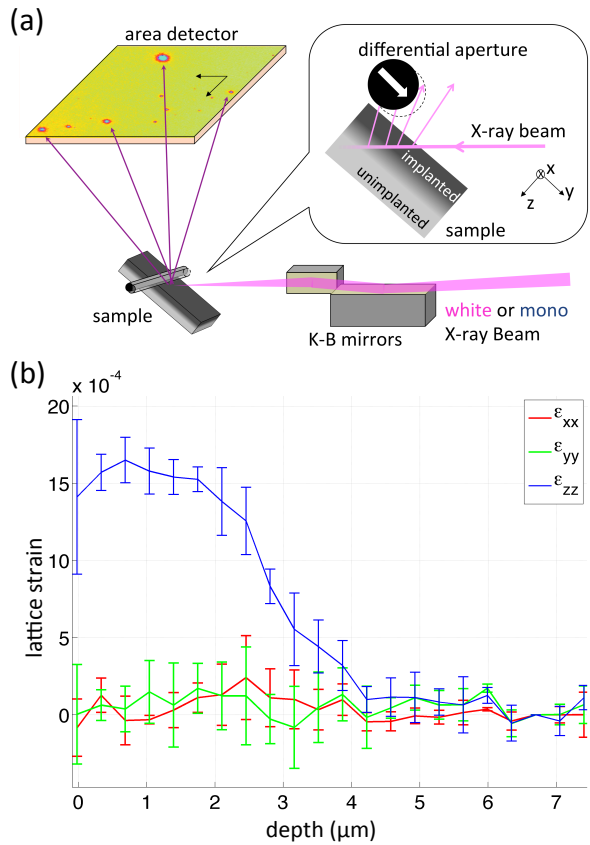

**Figure 2:** Laue diffraction measurements: (a) Schematic of the experimental configuration illustrating the orientation of the sample with respect to the incident beam, scanning wire and detector, as well as the orientation of the sample coordinate axes. (b) Experimentally-measured direct elastic strains in sample coordinates. The error-bars indicate the variation associated with experimental measurements.

At depths between 1 and 2 µm the calculated helium implantation profile (Fig. 1 (a)) is comparatively flat with an average helium concentration of 3110 ± 270 appm. The measured average $\varepsilon_{zz}$ strain for these depths is $(1550 ± 120) \times 10^{-6}$. Since tungsten is almost perfectly elastically isotropic at room temperature [40-42], linear isotropic elasticity can be used to determine the hydrostatic strain, $\varepsilon_v$, due to helium-implantation-induced defects. Assuming a spatially homogeneous distribution of randomly-oriented defects in the implanted layer and given that $\varepsilon_{xx} = \varepsilon_{yy} = 0$, the $\varepsilon_{zz}$ lattice strain can be written as:

$$\varepsilon_{zz} = \frac{\varepsilon_v}{3} + 2\frac{v}{(1-v)}\frac{\varepsilon_v}{3}, \tag{1}$$

where $v$ is the Poisson ratio. The first term in Eqn. (1) captures the increase in $\varepsilon_{zz}$ due to the homogeneous hydrostatic strain, $\varepsilon_v$, i.e. lattice swelling. The second term is due to the Poisson effect which arises from the lateral constraint on the implanted layer, i.e. $\varepsilon_{xx} = \varepsilon_{yy} = 0$. From the measured $\varepsilon_{zz}$ we estimate $\varepsilon_v = 2620 \times 10^{-6} \pm 200 \times 10^{-6}$ (using $v = 0.28$ [40-42]).

The stresses in the implanted layer are given by the following expressions:

$$\sigma_{xx} = \sigma_{yy} = \frac{vE}{(1+v)(1-2v)}\varepsilon_{zz} = \frac{vE}{3(1-v)(1-2v)}\varepsilon_v, \tag{2}$$

$$\sigma_{zz} = \frac{(1-v)E}{(1+v)(1-2v)}\varepsilon_{zz} = \frac{E}{3(1-2v)}\varepsilon_v = K\varepsilon_v, \tag{3}$$

where $K$ is the bulk modulus and E the Young modulus. Assuming $v = 0.28$ and $E = 410$ GPa [40-42], we find $\sigma_{zz} = 0.81 \pm 0.06$ GPa and $\sigma_{xx} = \sigma_{yy} = 0.32 \pm 0.02$ GPa. The $\sigma_{zz}$ stress component is associated with the density of internal forces exerted on the material by the helium-implantation-induced defects homogeneously distributed in the implanted layer. The $\sigma_{xx}$ and $\sigma_{yy}$ components are due to the same internal defect-induced forces *and* the boundary condition requiring the continuity of the material at the interface between the implanted layer and the substrate. We note that the defect-induced stresses in the implanted surface layer represent a significant fraction of the yield stress of the material, which is close to 1 GPa.

The stress contribution, $\sigma_{ii}^{BC}$, due to the boundary conditions in our sample is given by:

$$\sigma_{ii}^{BC} = \sigma_{ii} - p^{defect}, \tag{4}$$

where $p^{defect}$ is the hydrostatic stress due to internal defect forces. Since $\sigma_{zz}^{BC}$ = 0 GPa, it follows that $p^{defect}$ = 0.81 ± 0.06 GPa and hence $\sigma_{xx}^{BC} = \sigma_{yy}^{BC}$ = -0.49 ± 0.08 GPa.

2.3. **Surface acoustic wave measurements:**

Measurement of the Rayleigh velocity, $c_r$, of surface acoustic waves (SAW) provides an accurate means of determining the elastic properties of the helium-implanted layer. We use the laser-induced transient grating (TG) technique to measure the change of $c_r$ between implanted and unimplanted samples. A schematic of the TG setup, described elsewhere in more detail [43], is shown in Fig. 3 (a). Two short excitation pulses (515 nm wavelength, 60 ps pulse duration and 1.75 μJ pulse energy) are crossed on the sample. Interference of the two pulses results in a spatially sinusoidal excitation pattern, which, when absorbed, causes rapid thermal expansion and generation of two counter propagating SAWs. The SAW wavelength, λ, is determined by the period of the interference pattern. Here a nominal value of λ = 2.75 μm was chosen to ensure that the SAW signal in the helium-implanted sample is dominated by the ~2.8 μm thick implanted surface layer [44].

Detection of the SAW is achieved by diffraction of a quasi-continuous wave probe beam (532 nm wavelength and 10 mW average power) from the induced sample surface displacement and refractive index variations. The diffracted beam is collinear with a reflected reference beam to achieve heterodyne detection (Fig. 3 (a)). The combined beam is sent to a fast avalanche photo-diode (1 GHz bandwidth) and time traces are recorded with an oscilloscope. The excitation spot measured 500 μm diameter at $1/e^2$ intensity level and the probe spot 300 μm.

The recorded probe signal for the unimplanted W – 1%Re sample (Fig. 3 (b)) shows an exponentially decaying background, due to the decaying temperature grating, on which oscillations, due to the surface acoustic waves, are superimposed. By taking the Fourier transform of the time trace (inset in Fig. 3 (b)) the frequency of the Rayleigh waves, $f$, can be determined. An approximate solution for the Rayleigh wave speed, $c_r$, in an elastically isotropic medium is given by [45]:

$$c_r = f\lambda \approx \left(0.874 + 0.196\nu - 0.043\nu^2 - 0.055\nu^3\right)\sqrt{\frac{E}{2(1+\nu)\rho}}, \qquad (5)$$

where $\rho$ is the mass density (19260 kg m$^{-3}$ for the samples used here[40-42]). The integral error of this approximate solution over the range $\nu \in [\;-1 \quad 0.5\;]$ is 0.20% [46], which is sufficiently accurate for the relative comparisons performed here.

For the unimplanted W - 1 % Re sample we measured $c_r$ = 2680 ± 2 ms$^{-1}$ and for the helium-implanted sample $c_r$ = 2621 ± 7 ms$^{-1}$, i.e. $c_r$ is reduced by 2.2 %. The SAW measurements averaged over several grains in the sample and no significant grain-to-grain variation of $c_r$ was found, suggesting that tungsten remains close to being isotropically elastic after helium implantation. In a W - 1% Ta alloy implanted to the same helium dose under the same conditions a similar reduction of $c_r$ by 1.6 % was observed.

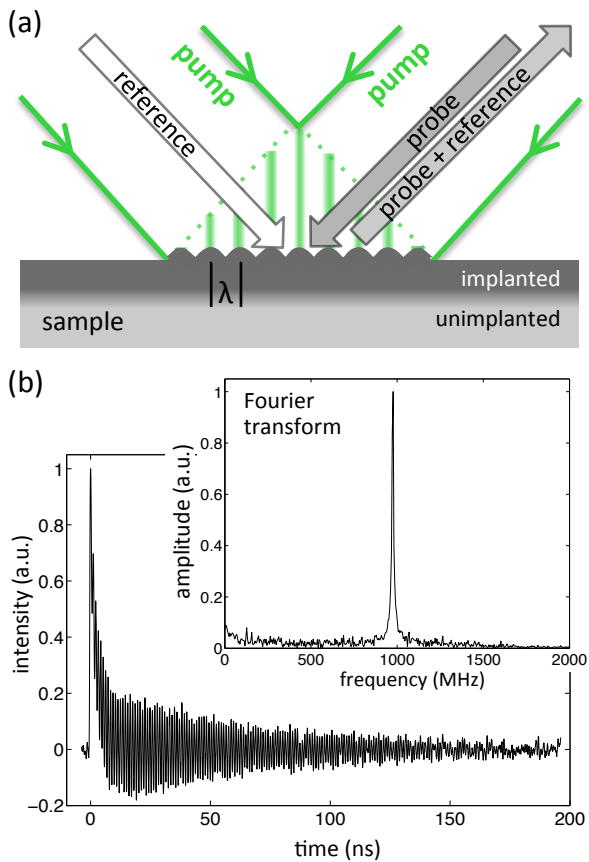

**Figure 3:** Transient grating surface acoustic wave measurements: (a) Schematic of the experimental setup. (b) Measured intensity signal plotted as a function of delay time relative to the pump pulse for the unimplanted W – 1 % Re sample. Fourier transform of the signal showing a peak at the surface acoustic wave frequency (inset).

## 3. Calculations

To elucidate the mechanisms that lead to the experimentally observed lattice swelling and Rayleigh wave velocity reduction, we combined Density Functional Theory (DFT), Molecular Dynamics (MD) and elasticity calculations.

DFT calculations were performed using the Perdew-Burke-Ernzerhof electron exchange-correlation functional within generalized gradient approximation (PBE-GGA). We used the projector augmented wave (PAW) pseudopotentials implemented in the Vienna Ab-initio Simulation Package (VASP) [47, 48]. Given that semicore electronic states make a non-negligible contribution to the formation energies of self-interstitial atom (SIA) defects [49], all the calculations were performed using the PAW potentials *Xpv*, where the semicore *p* states are treated as valence states. In all the cases considered here a 4 × 4 × 4 bcc supercell, with plane wave cut-off energy of 400 eV, and a 4 × 4 × 4 *k*-point mesh with spacing of 0.15 Å$^{-1}$ were used. Helium-induced defects, such as interstitial helium-clusters, helium-vacancy clusters and <111> SIA defects were introduced into the cell. The system was then fully relaxed, with unconstrained, periodic boundary conditions.

Larger-scale MD calculations were carried out using the most recent interatomic potential for body-centered cubic (bcc) tungsten [50]. Self-interstitial atom defects dominating swelling in the material (see the DFT section below) with random orientations and spatial distribution were introduced in an otherwise perfect bcc box containing 65536 atoms (32x32x32 bcc unit cells in the normal Cartesian coordinate system). To reproduce the lateral constraint matching that of experiments, the size of the simulation box was fixed in two directions (x = [100] and y = [010]). Unconstrained expansion was only allowed in the third direction ($z$=[001]). Molecular relaxation was then performed assuming periodic boundary conditions in all directions.

4. **Results and Discussion**

**4.1. Formation and relaxation volumes of defects**

Relating the observed macroscopic strains in the helium-implanted layer to the microscopic characteristics of individual defects requires evaluation the elastic dipole tensor of defects $P_{ij}$ [51], also known as the double-force tensor [52]. In cubic crystals the trace of a defect dipole tensor is proportional to the product of the bulk modulus, K, and the relaxation volume of the defect, $\Omega_r$ [53]:

Tr $P_{ij}$=3 K $\Omega_r$. (6)

Using DFT, we calculate $\Omega_r$ for a number of helium-implantation-induced defects. $\Omega_r$ is defined as the volume change associated with the introduction of a defect in a simulation cell, i.e.:

$$\Omega_r = \Omega(\text{defect}) - \Omega(\text{perfect}).$$ (7)

Here $\Omega$(perfect) is the volume initially occupied by the perfect, defect free 4x4x4 supercell, whereas $\Omega$(defect) is the volume occupied by the relaxed supercell containing a defect. Relaxation volume provides a measure of elastic strain in the vicinity of the defect [54]. It enters the elasticity equations for the stress and strain fields resulting from the accumulation of defects.

Tab. 1 lists $\Omega_r$ values calculated for a number of helium-implantation-induced defects. Tab. 1 (a) shows that $\Omega_r$ of vacancies and vacancy clusters is negative, whilst $\Omega_r$ for SIAs is large and positive. The relaxation volume of a Frenkel pair, $\Omega_r$(Frenkel), is given by the sum $\Omega_r$(Frenkel) = $\Omega_r$(V) + $\Omega_r$(SIA) = 1.31 $\Omega_0$. As expected, $\Omega_r$(Frenkel) is positive and greater than $\Omega_0$, the perfect atomic volume (15.85 Å$^3$ in tungsten at 298 K [38]) . Relaxation volumes for interstitial He$_n$ clusters (Tab. 1 (b)) are all positive and, to a good approximation, increase linearly with $n$ for $1 \leq n \leq 5$. The calculated relaxation volumes for vacancy defects and interstitial helium defects are in good agreement with values reported by other authors [55-57]. For He$_n$V defects (Tab. 1(c)) the relaxation volume changes sign from negative to positive as the number of helium atoms increases. For a He$_2$V defect the relaxation volume is close to zero.

(a) Relaxation volumes of vacancies and self-interstitial atom defects

| V | $V_2$ (1NN) | $V_2$ (2NN) | $V_2$(3NN) | $V_3$(1NN(2)+2NN) | <111> SIA |
|---|---|---|---|---|---|
| -0.37 | -0.72 | -0.79 | -0.76 | -1.08 | 1.68 |
| -0.34 [55] | -0.65 [55] | -0.74 [55] | -0.69 [55] | | |
| -0.38 [56] | | | | | |

(b) Relaxation volumes of interstitial helium clusters

| He (tetra) | He (octa) | $He_2$ (tetra) | $He_3$ (tetra) | $He_4$ (tetra) | $He_5$ (tetra) |
|---|---|---|---|---|---|
| 0.36 | 0.37 | 0.80 | 1.16 | 1.65 | 2.03 |
| 0.33 [57] | 0.34 [57] | | | | |

(c) Relaxation volumes of helium - vacancy clusters

| HeV (tetra) | HeV (octa) | $He_2$V (tetra) | $He_3$V (tetra) | $He_4$V (tetra) | $He_5$V (tetra) | $He_6$V (tetra) |
|---|---|---|---|---|---|---|
| -0.24 | -0.23 | -0.06 | 0.14 | 0.38 | 0.71 | 1.09 |

**Table 1:** Relaxation volumes, $\Omega_r$, of various defects in pure tungsten. Values are calculated by DFT and are given in atomic volume units $\Omega_0$. (a) Relaxation volumes of vacancies and self-interstitials (V: vacancy, $V_2$: di-vacancy, $V_3$: tri-vacancy, SIA: self-interstitial atom). (b) Relaxation volumes of interstitial $He_n$ clusters. (c) Relaxation volumes of $He_n$V clusters.

## 4.2. Lattice strain and swelling

Using the relaxation volumes of defects derived from DFT it is now possible to compute the lattice strain associated with the accumulation of helium and implantation-induced defects in the implanted surface layer. It is essential to distinguish between two possible modes of swelling: One is associated with the accumulation of vacancies in the bulk of the material and migration of self-interstitial atoms to the surface. This (Shottky) swelling mode does explain the increase of volume of the sample, but predicts very low elastic strain. Moreover, since the relaxation volume of vacancy defects is negative, the lattice strain associated with this mode of swelling is expected to be compressive (negative). This is not observed in our experiments. The alternative (Frenkel) model assumes that both self-interstitial and vacancy defects are retained in the implanted layer, and that the observed strains and stresses result from the forces exerted by the defects on the surrounding material. The experimental X-ray diffraction data suggest that it is the Frenkel model for defect accumulation and lattice swelling that describes our implantation experiments.

To evaluate lattice swelling associated with defects accumulated in the implanted layer, we consider the density of forces exerted on the material by the defects [51]:

$$F_i(\mathbf{r}) = -\sum_\alpha P_{il}^{(\alpha)} \frac{\partial}{\partial r_l} \delta(\mathbf{r} - \mathbf{R}_\alpha), \tag{8}$$

where $\mathbf{R}_\alpha$ is the coordinate of a defect and $\mathbf{r}$ the coordinate vector. The stress tensor can now be evaluated by considering the equilibrium condition [51]:

$$\frac{\partial}{\partial r_k} \sigma_{ik} + F_i = 0. \tag{9}$$

If an implanted volume is free of constraints in all directions, averaging over the distribution of defects and random orientations of defects gives:

$$\sigma_{ii} = p^{defect} = \frac{1}{3}\sum_A n_A \text{Tr} P_{ij}^{(A)}, \tag{10}$$

where summation is performed over defect types, A, each with number density $n_A$, and $\sigma_{ii}$ is as described by Eqn. (4) for the case of $\sigma_{ii}^{BC} = 0$. Using Eqn. (6), we arrive at the following expression for defect-induced lattice swelling:

$$\varepsilon_v = \sum_A n_A \Omega_r^{(A)}. \tag{11}$$

This equation is one of the central results of this paper. It relates the macroscopic lattice swelling to the average concentrations of defects and their relaxation volumes. Combining it with Eqn. (1), we arrive at an expression for the out-of-plane lattice strain, $\varepsilon_{zz}$, that is expected for a given defect population in an implanted layer subject to the lateral constraints present in our sample:

$$\varepsilon_{zz} = \frac{1}{3}\frac{(1+\nu)}{(1-\nu)}\sum_A n_A \Omega_{rel}^{(A)}. \tag{12}$$

Using this equation, we can now quantitatively study the type of microstructure formed as a result of helium implantation. *Ab-initio* studies point out that helium has a strong affinity to vacancies [18]. Hence we first assume that each implanted helium atom occupies a vacancy, preventing recombination of some of the vacancies and SIAs formed during implantation. Using Eqn. (12) and the relaxation volumes given in Table 1, the strain due to a homogeneous distribution of 3110 appm of HeV defects and the corresponding 3110 appm of SIAs can be evaluated. Such an "evenly balanced" assumed microstructure is characterised by an average strain $\varepsilon_{zz}$ = 2654 x $10^{-6}$. This is almost twice the level of strain observed experimentally.

Helium clustering offers an explanation for the lower experimentally observed $\varepsilon_{zz}$ strain. According to our DFT calculations, which agree with the analysis performed earlier [20, 58], vacancies can be readily occupied by several helium atoms. This reduces the concentration of <111> SIA defects and the strain in the implanted layer. If we assume that all the vacancies are occupied by two helium atoms (i.e. that the defect population consists of 1555 appm $He_2V$ and 1555 appm SIAs), we predict the out of plane elastic strain $\varepsilon_{zz}$=1493 x $10^{-6}$. This agrees very well with the measured out of plane strain of $\varepsilon_{zz}$ = 1550 x $10^{-6}$.

From DFT it is also possible to evaluate the formation energy, $E_f$, associated with the formation of $He_nV$ clusters. This can be normalised by the number of helium atoms, $n$, to give: $\bar{E}_f(He_nV) = E_f(He_nV)/n$. Fig. 4 shows a plot of the calculated $\bar{E}_f(He_nV)$ values as a function of $n$. $\bar{E}_f(HeV)$ is more than 1 eV greater than $\bar{E}_f(He_2V)$, whilst there is little change in $\bar{E}_f(He_nV)$ for $n$ in the range from 2 to 5. This indicates that from an energetic point of view the formation of $He_nV$ clusters containing more than one helium atom is favourable, in agreement with our swelling analysis.

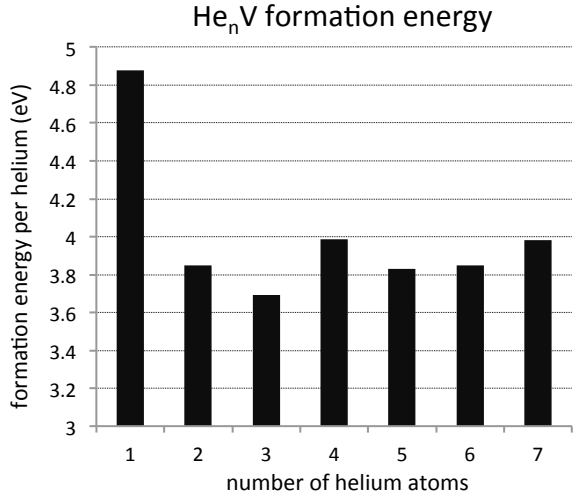

**Figure 4:** $\bar{E}_f(\text{He}_n\text{V})$ plotted as a function of *n*. $\bar{E}_f(\text{He}_n\text{V})$ is the formation energy of $\text{He}_n\text{V}$ defects calculated by DFT, normalised by the number of helium atoms in a vacancy, *n*.

In experiments we measure elastic strain, an integral quantity to which the entire population of helium and radiation-induced defects contribute. Because of this we cannot be more precise in our assessment of defect microstructure formed as a result of helium implantation. Nevertheless, we are able to conclude without ambiguity that helium implanted in tungsten at 300°C forms helium-vacancy clusters that *predominantly contain more than one helium atom per vacancy*. Furthermore, as detailed below, we find that the observed lattice swelling is almost entirely due to the accumulation of self-interstitial atom defects, characterised by large relaxation volumes, which cause lattice expansion.

Two key assumptions in our defect analysis are: first that the orientations of the defects are random; and second that swelling scales linearly with defect concentration, which is the case in the low concentration limit. To validate these assumptions we performed a series of larger-scale MD simulations. From DFT calculations we found that the measured lattice swelling is consistent with a defect structure dominated by $He_2V$ clusters and SIAs. The relaxation volumes of a $He_2V$ cluster and a <111> SIAs are -0.06 $\Omega_0$ and 1.68 $\Omega_0$ respectively (Tab. 1). This means that $He_2V$ clusters contribute little to lattice volume change. Lattice swelling is instead predominantly due to SIAs. Thus in our MD calculations only SIAs were taken into account. The effective SIA concentration was varied from 1000 to 4000 appm, and for each case 10 random SIA configurations were relaxed. Fig. 5 a) and b) respectively show the calculated out of plane strain ($\varepsilon_{zz}$) and the in-plane residual stresses due to the lateral constraints ($\sigma_{xx}^{BC} = \sigma_{yy}^{BC}$) as a function of SIA concentration. Both quantities show a near perfect linear dependence on concentrations, confirming the appropriateness of the low concentration limit assumption. For an SIA concentration of 1555 appm the predicted out of plane strain, $\varepsilon_{zz}$, is 1240 x $10^{-6}$ compared to a measured value of 1550 x $10^{-6}$. The magnitude of the predicted in-plane residual stresses due to boundary conditions, $\sigma_{xx}^{BC} = \sigma_{yy}^{BC}$ is -0.39 GPa, somewhat lower that the value of -0.49 GPa determined from experiments. Thus it appears that MD calculations underestimate the helium-implantation-induced stresses and strains by 20 %. Still, this is reasonable agreement given the absence of fitting parameters other than those involved in the fitting of the interatomic potential.

Fig. 5 (c) shows a relaxed configuration for an SIA concentration of 1602 appm, corresponding to 105 SIAs in the simulation box. Delocalisation of SIAs into crowdions is clearly visible [59]. They are aligned along a number of different <111> directions, justifying the assumption of a randomly oriented defect population.

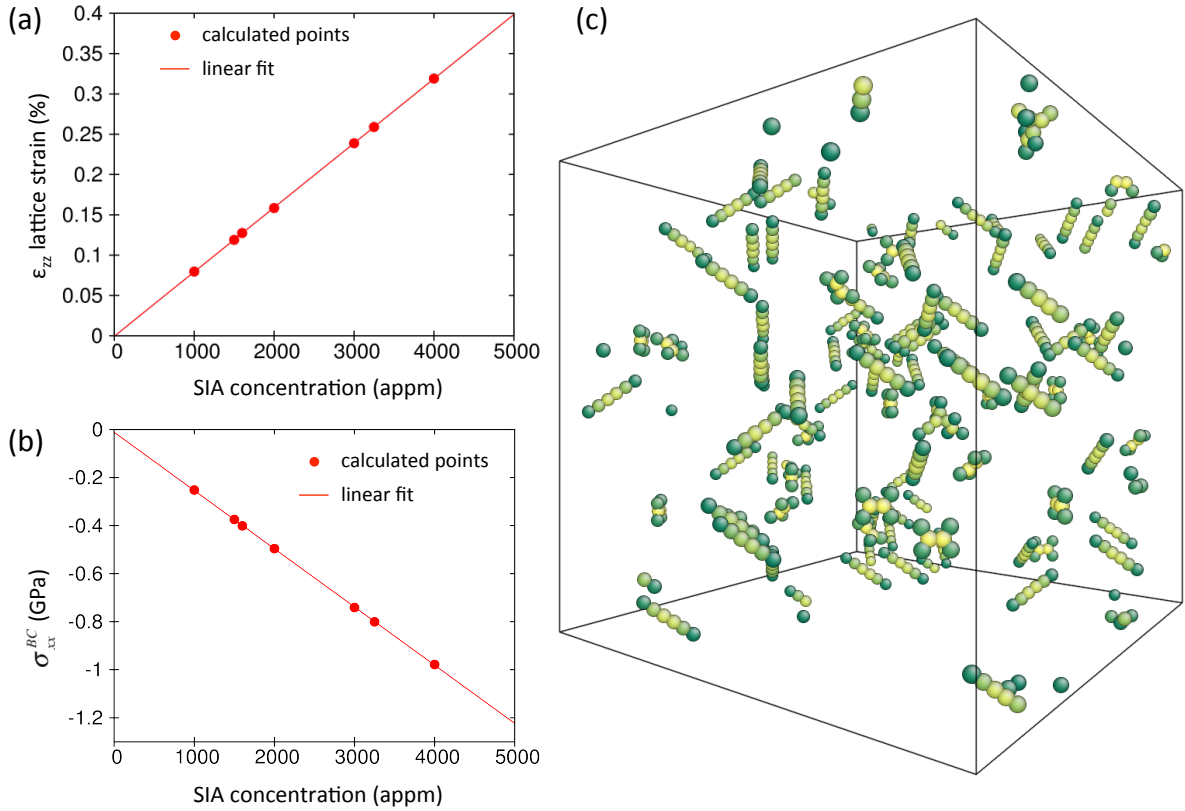

**Figure 5:** MD calculation results for lattice swelling and stresses due to SIAs. (a) Calculated $\varepsilon_{zz}$ lattice strain as a function of SIA concentration. (b) $\sigma_{xx}^{BC}$ stress, due to lateral boundary conditions, as a function of SIA concentration. (c) Relaxed configuration snapshot of MD calculation with 105 SIA defects (1602 appm SIA concentration).

We note in passing that our analysis also indicates that the vast majority of Frenkel defects, generated during helium implantation, recombine. Only a small proportion is being prevented from doing so by the injected helium. According to our SRIM calculations, each implanted helium ion generates on average ~74 Frenkel pairs (see Fig. 1 (b) and (c)). Of these, only 0.5 Frenkel pairs are prevented from recombination. This means that less than 1% of the calculated dpa is actually retained in the material, making the value of using dpa as a measure of the "actual" irradiation-induced damage questionable.

### 4.3. Surface acoustic wave velocity reduction

To elucidate the decrease in the surface acoustic wave velocity resulting from helium implantation, the changes in elastic properties that arise due to He$_2$V clusters and <111> SIAs were evaluated using DFT. A detailed description of the method used to calculate elastic constants is given elsewhere [60-63].

The fully relaxed 4x4x4 DFT supercell for perfect tungsten exhibits the expected O$_h$ symmetry. The calculated elastic constants $C_{11}$, $C_{12}$ and $C_{44}$ are listed in Tab. 2 (a). When an SIA is introduced into the simulation cell, a [111] crowdion is formed [49] (see relaxed configuration shown in Fig. 6 (a)). The relaxed 4x4x4 supercell containing one SIA exhibits a significant distortion and the resulting loss of symmetry. The supercell now has D$_{3d}$ symmetry, retaining 3 reflection and 3 rotation symmetries of the total of 48 symmetries of the original defect-free structure. This means that 8 distinct defect orientations exist. To obtain the elastic constants of a 4x4x4 supercell containing a randomly orientated SIA defect the stiffness tensor calculated by DFT was averaged over all 8 possible configurations. The resulting constants are listed in Tab. 2 (a).

For the He$_2$V cluster a vacancy was introduced in the centre of the 4x4x4 supercell and two helium atoms were placed in the tetrahedral positions at opposite sides of the vacancy. Fig. 6 (b) shows the He$_2$V cluster after relaxation. The two helium atoms align along the [110] direction and form a defect in the ($\bar{1}$10) plane. The resulting supercell has C$_{1h}$ symmetry, retaining only one reflection symmetry. Hence there are 24 distinct defect orientations. The elastic constants, found for the 4x4x4 supercell when averaging over these configurations, are listed in Tab. 2 (a).

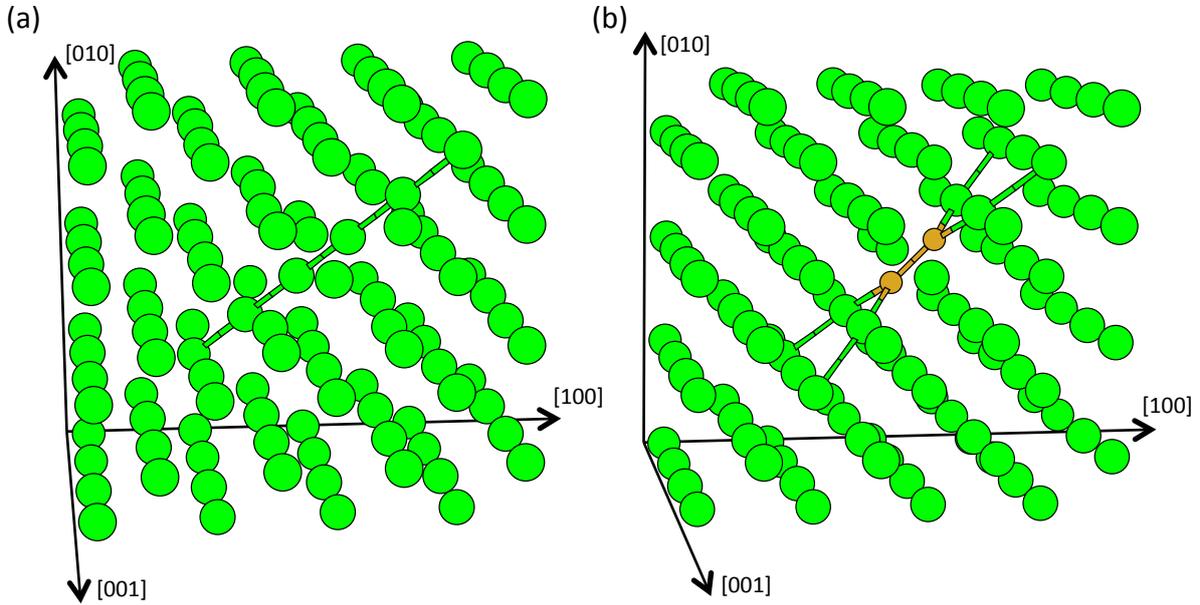

**Figure 6:** Relaxed defect configurations from DFT for: (a) [111] SIA. (b) He$_2$V cluster.

The elastic constants calculated by DFT for defect-free tungsten in a 4x4x4 supercell (Tab. 2(a)) correspond to the anisotropy factor $A = 2C_{44}/(C_{11} - C_{12}) = 0.88$. Experimentally measured elastic constants of tungsten show a decrease of A from A = 1.01 at room temperature (T = 298 K) [40-42] to A = 0.98 at T = 4.2 K [64], whilst also indicating that tungsten is close to being elastically isotropic. The DFT value for the anisotropy factor, calculated at T=0K, reproduces the experimental trend of A<1 at low temperature, but its value deviates from the experimentally measured one. There are two main factors that explain this difference: The first is that elastic constants were evaluated at the DFT equilibrium volume. This is often slightly different from the experimental one; the difference depends on the choice of exchange-correlation functional. The second is the sensitivity of the computed elastic constants to k-space sampling. We also performed DFT calculations for the cases of perfect tungsten with 2 atoms per bcc cubic cell and 1 atom per bcc primitive cell. The smaller cell size allowed for a finer k-space sampling. The calculated values of A were 0.89 and 0.93, for 2 atoms and 1 atom respectively, closer to experimental observations.

To correct for the difference between experiments and DFT calculations we introduce rescaling factors $f_{11}$, $f_{12}$ and $f_{44}$, defined as $f_{ij} = C_{ij}^{\text{exp}}(\text{pure W}) / C_{ij}^{DFT}(\text{no defect})$. Here $C_{ij}^{\text{exp}}(\text{pure W})$ are the experimentally measured constants for pure tungsten at room temperature (Tab. 2 (a)) and $C_{ij}^{DFT}(\text{no defect})$ are the constants calculated from DFT for the defect-free 4x4x4 tungsten supercell. Treating the changes to elastic constants that arise due to He$_2$V and SIA defects as small perturbations, the same rescaling factors were applied to the elastic constants calculated by DFT for a supercell containing He$_2$V or SIA defects, i.e.:

$$C_{ij}^{\text{DFT rescaled}}(\text{defect}) = f_{ij} C_{ij}^{\text{DFT}}(\text{defect}). \tag{13}$$

The resulting values for elastic constants are listed in Tab. 2 (a). The elastic constants for the helium-implanted material were then calculated using the Voigt approach [65]:

$$C_{ij}^{implanted} = (1 - 128(n_{SIA} + n_{He_2V}))C_{ij}^{W} + 128 n_{SIA} C_{ij}^{SIA} + 128 n_{He_2V} C_{ij}^{He_2V}, \tag{14}$$

where $n$ are the number densities of SIA and He$_2$V defects (1555 appm), and the factor of 128 accounts for the number of atoms contained within the DFT supercell for which elastic constants were computed. The resulting elastic constants for the helium-implanted layer are given in Tab. 2 (b).

For helium-implanted tungsten A remains close to 1, such that it may be reasonably approximated isotropic elasticity. The bulk modulus, K, and shear modulus G are given by [65]:

$$K = \frac{1}{3}(C_{11} + 2C_{12}), \qquad (15)$$

$$G = \frac{1}{5}(C_{11} - C_{12} + 3C_{44}). \qquad (16)$$

Using isotropic elasticity relations, it is then a straightforward matter to calculate the Young modulus, E, and the Poisson ratio, $\nu$:

$$E = \frac{9K}{1 + 3\frac{K}{G}}, \qquad \nu = \frac{\left(K - \frac{2}{3}G\right)}{2\left(K + \frac{2}{3}G\right)}. \qquad (17)$$

Using Eqn. (5) we evaluate the Rayleigh wave velocity, $c_r$. For unimplanted tungsten we find $c_r$ = 2666 ms$^{-1}$, in good agreement with the experimental value of 2680 ± 2 ms$^{-1}$. The predicted decrease of the Rayleigh velocity following helium-implantation is 1.7 %, compared to the experimentally measured decrease of 2.2%. This agreement is remarkable since, other than the scaling factors of Eqn. (13), which in any case do not influence the *ratios* of the relevant elastic constants, there are no adjustable parameters in our model. Furthermore we note that this reduction can only be achieved by considering *both* the SIA and He$_2$V cluster contributions. Separately, SIAs and He$_2$V clusters would only yield reductions of 1.0 % and 0.7% respectively.

Similar calculations were also carried out using the Reuss approach [66]. The calculated $c_r$ values for pure tungsten and helium-implanted tungsten are 2666 ms$^{-1}$ and 2617 ms$^{-1}$ respectively, indicating a reduction of 1.8%. This is very similar to the value calculated using the Voigt approach, as expected, given that the material remains nearly elastically isotropic after implantation.

(a) Elastic properties calculated by DFT

|  |  | $C_{11}$ (GPa) | $C_{12}$ (GPa) | $C_{44}$ (GPa) | A |
|---|---|---|---|---|---|
| Pure W, experiment at 298 K [40-42] | | 522.8 | 203.5 | 160.7 | 1.01 |
| Pure W | from DFT | 537.4 | 188.2 | 153.7 | 0.88 |
| | rescaled | 522.8 | 203.5 | 160.7 | 1.01 |
| W + SIA | from DFT | 512.5 | 212.5 | 141.4 | 0.94 |
| | rescaled | 498.6 | 229.8 | 147.8 | 1.10 |
| W + $He_2V$ | from DFT | 518.9 | 188.1 | 141.2 | 0.85 |
| | rescaled | 504.8 | 203.4 | 147.6 | 0.98 |

(b) Calculated elastic constants for the helium-implanted layer

|  |  | $C_{11}$ (GPa) | $C_{12}$ (GPa) | $C_{44}$ (GPa) | A |
|---|---|---|---|---|---|
| Helium-implanted W | from DFT | 528.8 | 193.0 | 148.8 | 0.89 |
| | rescaled | 514.4 | 208.7 | 155.5 | 1.02 |

(c) Isotropic elastic constants (calculated using Voigt) and surface acoustic wave velocity

|  |  | K (GPa) | G (GPa) | E (GPa) | nu | $c_r$ (ms$^{-1}$) |
|---|---|---|---|---|---|---|
| Pure W | from DFT | 304.6 | 162.1 | 413.0 | 0.274 | 2679 |
| | rescaled | 309.9 | 160.3 | 410.1 | 0.279 | 2666 |
| Helium-implanted W | from DFT | 304.9 | 156.4 | 400.7 | 0.281 | 2634 |
| | rescaled | 310.6 | 154.5 | 397.5 | 0.287 | 2621 |

**Table 2:** (a) Elastic parameters of perfect tungsten and tungsten containing SIA and $He_2V$ defects calculated by DFT for a 128 atom simulation cell. Both as-calculated and rescaled values (highlighted in grey) are listed. (b) Calculated elastic constants for helium-implanted tungsten (W + 1555 appm $He_2V$ + 1555 appm SIAs). (c) Isotropic elastic constants and SAW velocities for pure tungsten and helium-implanted tungsten. Values calculated based on rescaled elastic constants are highlighted in grey.

(a) Elastic constants calculated by MD using the Marinica et al. potential [50]

|  | $C_{11}$ (GPa) | $C_{12}$ (GPa) | $C_{44}$ (GPa) | A |
|---|---|---|---|---|
| Pure W | 523 | 202 | 161 | 1.00 |
| W + 1555 appm SIAs | 528.8 | 208.7 | 168.0 | 1.05 |

(b) Isotropic elastic constants (calculated using Voigt) and surface acoustic wave velocities

|  | K (GPa) | G (GPa) | E (GPa) | Nu | $c_r$ (ms$^{-1}$) |
|---|---|---|---|---|---|
| Pure W | 309.0 | 160.8 | 411.1 | 0.278 | 2670 |
| W + 1555 appm SIAs | 315.4 | 164.8 | 421.1 | 0.278 | 2703 |

**Table 3:** (a) Elastic properties of defect-free tungsten and tungsten containing 1556 appm SIAs calculated from MD. (b) Isotropic elastic constants and surface acoustic wave velocities computed for defect free tungsten and tungsten containing 1556 appm SIAs.

We also evaluated the elastic constants predicted by MD calculations used for the lattice swelling analysis, but this time without the lateral constraint during relaxation, which is consistent with the DFT calculations. The resulting constants are listed in Tab. 3 (a) for defect free tungsten and a 32 x 32 x 32 unit cell simulation box containing 102 randomly oriented SIAs (i.e. 1556 appm SIA concentration). The Voigt isotropic elastic constants were then calculated using equations (15-17) and the surface acoustic wave velocities estimated using equation (5) (Tab. 3 (b)). We note that MD calculations predict an *increase* of surface acoustic wave velocity by 1.2 % due to SIAs. This is not in agreement with either our DFT calculations, or the experimental measurements. The most probable reason for the disagreement is that the interatomic potential used in MD simulations is not able to correctly describe the variation of elastic constants as a function of strain, a point that is fairly difficult to include in the existing potential fitting methodology [50].

## 4. Conclusions:

In this paper we have demonstrated the feasibility of using synchrotron X-ray micro-beam diffraction to quantify lattice swelling in a few micron-thick helium-implanted tungsten layer. Using elasticity theory and relaxation volumes calculated by DFT for various helium-induced defects, we explained the experimentally measured swelling. Agreement between calculations and experiments is excellent, confirming that small helium-vacancy clusters and SIAs dominate the helium-implantation-induced microstructure. MD calculations considering the effect of SIAs alone agreed well with the experimentally observed swelling, indicating that lattice swelling is indeed dominated by the presence of SIAs. Surface acoustic wave measurements, using the transient grating technique, showed a decrease of Rayleigh velocity in the helium-implanted material. Using a model based on DFT and elasticity theory, we interpret the reduction in Rayleigh wave velocity, in very good agreement with experiments.

## 6. Acknowledgements:

We thank B. Abbey, A. De Backer and S. G. Roberts for helpful comments and stimulating discussions, as well as A. Xu and G. Hughes for help with sample preparation. FH acknowledges funding from the John Fell fund (122/643) and the Royal Society (RG130308). DNM acknowledges the International Fusion Energy Research Centre (IFERC) for use of the supercomputer (Helios) at the Computational Simulation Centre (CSC) in Rokkasho (Japan). DEJA acknowledges the Royal Academy of Engineering for support through a Research Fellowship. The SAW measurements at MIT were supported by NSF Grant no. CHE-1111557. Use of the Advanced Photon Source, an Office of Science User Facility operated for the U.S. Department of Energy (DOE) Office of Science by Argonne National Laboratory, was supported by the U.S. DOE under Contract No. DE-AC02-06CH11357. This work was also part-funded by the RCUK Energy Programme (Grant Number EP/I501045) and by the European Union Horizon 2020 research and innovation program grant number 633053. To obtain further information on the data and models underlying this paper please contact PublicationsManager@ccfe.ac.uk. The views and opinions expressed herein do not necessarily reflect those of the European Commission. This work was part-funded by the United Kingdom Engineering and Physical Sciences Research Council via programme grants EP/G050031 and EP/H018921.